\documentclass[twocolumn,notitlepage,pra,superscriptaddress]{revtex4-1}
\usepackage{graphicx}
\usepackage{braket}
\usepackage{amsmath,amsfonts,amssymb,amstext,amscd,amsthm,bbold}
\usepackage{comment,float}
\usepackage{tabularx}
\usepackage{tikz}
\usetikzlibrary{quantikz}

\usepackage[colorlinks=true,linkcolor=blue,urlcolor=magenta,citecolor=magenta]{hyperref}
\usepackage[normalem]{ulem}
\usepackage{dsfont}
\usepackage{braket}
\usepackage{rotating}
\usepackage{subfig}
\usepackage{lipsum}
\usepackage{CJK}
\usepackage{multirow}
\usepackage{lipsum}

\begin{document}
	\title{Deterministic quantum teleportation of a path-encoded state using entangled photons}
	
	\author{Akshai T. Krishnan}
	\affiliation{Dept of Instrumentation and Applied Physics, Indian Institute Science, Bengaluru 560012, India}
	\author{Kanad Sengupta }
	\affiliation{Dept of Instrumentation and Applied Physics, Indian Institute Science, Bengaluru 560012, India}
	\author{S. P. Dinesh}
	\affiliation{Dept of Instrumentation and Applied Physics, Indian Institute Science, Bengaluru 560012, India}
	\author{C. M. Chandrashekar}
	\affiliation{Dept of Instrumentation and Applied Physics, Indian Institute Science, Bengaluru 560012, India}
	\affiliation{The Institute of Mathematical Sciences, C. I. T. Campus, Taramani, Chennai 600113, India}
	\affiliation{Homi Bhabha National Institute, Training School Complex, Anushakti Nagar, Mumbai 400094, India}

	
\begin{abstract}
 Quantum teleportation enables a way to transmit an arbitrary qubit state from one place to an other. A standard scheme for teleportation in optical setup involve three photons,  an entangled photon pair and a photon carrying quantum state to be teleported. The interaction between the photons in the scheme makes  quantum teleportation probabilistic. Here we demonstrate a deterministic teleportation of an arbitrary qubit state using only a polarization entangled photon pair in a linear optical setup. By introducing the path degree of freedom to one of the entangled photon and encoding the arbitrary qubit state into it, we demonstrate  100\% Bell State Measurement (BSM) outcome. This enables a deterministic teleportation of a qubit state in an optical scheme with high fidelity. We report an average teleportation fidelity of 88.00$\pm$0.04\%. The dependency of fidelity on the visibility of single-photon interferometer used for path qubit state shows the possibility of further improving the fidelity of quantum teleportation.
\end{abstract}
	
\maketitle    

{\it{Introduction:}} 
Quantum teleportation enables the reconstruction of arbitrary quantum state between two spatially separated particles without actually transmitting the quantum state or a particle transfer\,\cite{CGCA1993,SA1995,DJKM1997}. Using a shared entangled state between two parties and a classical communication channel, quantum teleportation has been experimentally demonstrated in various quantum systems\,\cite{GS1999,TKCH2003,PHXC2016,PP2021,SSH2002,HSA2007,AM1998,YSKY2001,DEKY2004,JHL2023,YSXB2008,NAC2013}.
 Among them, the ability to share entangled photons over large distances is poised to play an important role as a resource for quantum communication, distributed quantum networks and quantum computation\,\cite{SJCA2015,XYCG2023,MMR2009}. 
This has resulted in the demonstration of quantum teleportation of photonic qubit states encoded in different degrees of freedom of photons\,\cite{XXZM2015} and in higher dimensional states\,\cite{YHM2019,XCBY2020}.  However, even to teleport a standard two-level quantum state in a photonic system, one needs to consider a photon in an arbitrary quantum state encoded with Alice and its interaction with one of the entangled photon pairs which are shared between Alice and Bob. The interaction between the two photons with Alice is always  probabilistic in nature reducing the success rate of the Bell-state measurement (BSM) which generates classical information to be exchanged with Bob. However, by using different degrees of freedom of the photon to encode arbitrary qubit state, the best efficiency reported for performing BSM and teleport a qubit state is 50\%\,\cite{QYSW2016}. Apart from the qubit state, that is, the discrete variable approach, quantum teleportation has also been demonstrated in the continuous variable approach\,\cite{SH1998,HTT2000,PS2000,HTA2004,HIWH2014,WNBR2003} and hybrid configuration\,\cite{DAUE2018,GTHV2020}. There are also schemes available for multipartite quantum teleportation in both theory and experiment\,\cite{X2001, SWJ2003, G2005}.

Here we demonstrate the deterministic teleportation of qubit state encoded in spatial (path) mode using only photon pairs entangled in polarization degree of freedom. Encoding arbitrary qubit state in path degree of freedom on the entangled photon with Alice does not invoke any photon-photon interaction at Alice side. This enables a deterministic and controlled operations on qubits with Alice resulting in deterministic teleportation of arbitrary qubit state from Alice to Bob. Earlier reported works employing path encoded qubit states involve higher dimensional state teleportation with a best fidelity of 0.75 using probabilistic BSM\,\cite{XCBY2020,YHM2019} unlike the scheme reported here which demonstrate a deterministic teleportation of only two-dimensional state. 
\begin{figure}[H]
        \centering
        \includegraphics[width = 0.5\textwidth]{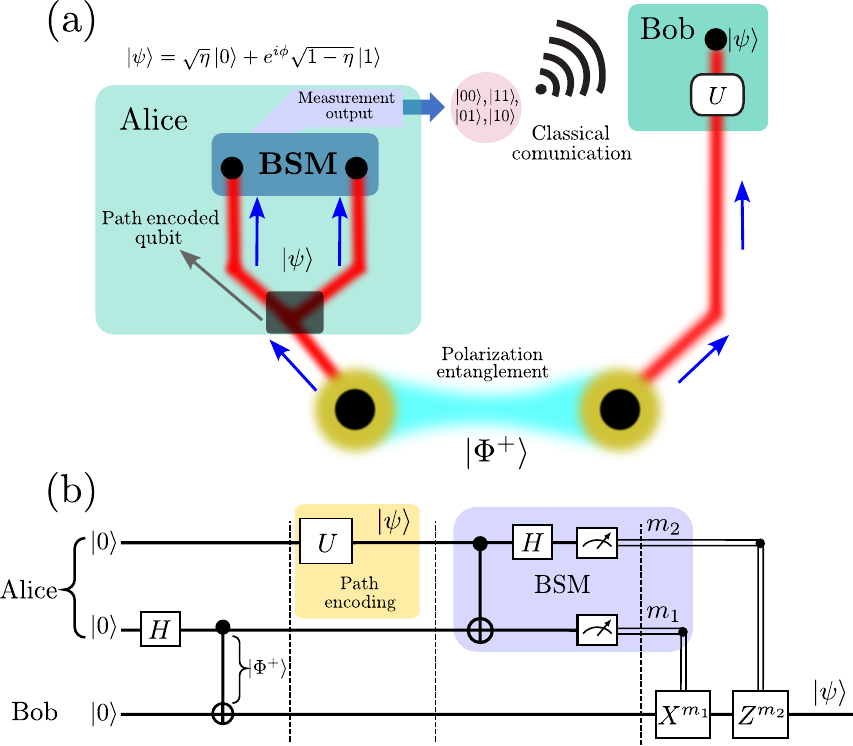}
        \caption{(a) Scheme for quantum teleportation of the path-encoded quantum state \(\ket{\psi}\) using a two-photon polarization-entangled state. Alice encodes the quantum state to be transferred in the path encoding and performs the Bell-state measurement (BSM) in two degrees of freedom: polarization and path. She then sends the relevant classical bits to Bob, who reconstructs the quantum state.   (b) This schematic shows the relevant circuit model for the teleportation scheme. The yellow portion represents the preparation of an arbitrary quantum state on Alice's side, while the magenta portion illustrates the BSM and classical communication.}
        \label{fig1}
\end{figure}

We report an average teleportation fidelity of 88.00$\pm$0.04\% with highest being 99.00\% for state $|0\rangle$ and 80.30\% for state $|R\rangle = (|0\rangle + i |1\rangle)/\sqrt{2})$. The fidelity of teleportation is dependent on the interference visibility of the single photon interference employed for BSM in the scheme.  With the increase in single photon interference visibility near cent percent teleportation can be achieved. In quantum communication protocols for teleportation of quantum states or for teleportation of qubit state in photonic quantum computation involving path degree of freedom\,\cite{MASK2024,KSKS2024}, the deterministic scheme presented here will be more easily employable when compared to other known approaches and can be extended to long distance quantum teleportation. 

\onecolumngrid	

\begin{figure}[H]
       \centering
        {\includegraphics[width = 0.95\textwidth]{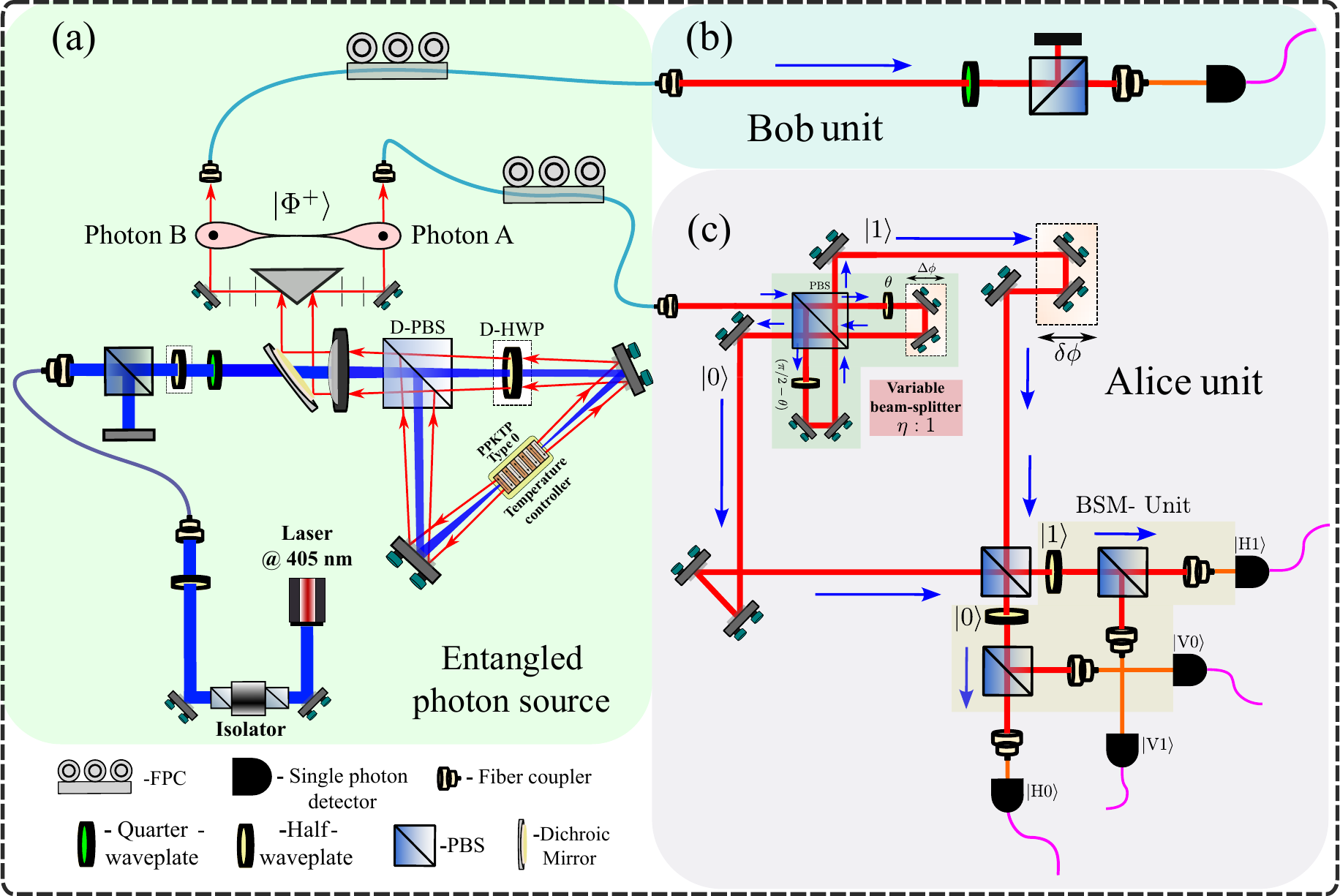}}
        \caption{Experimental setup for teleporting a qubit state using the spatial mode of a polarization-entangled photon, which itself acts as the teleportation quantum channel. (a) A continuous-wave laser with a wavelength of 405 nm is focused on a PPKTP crystal, utilizing type-0 phase matching to produce degenerate polarization-entangled photon pairs at 810 nm in a Sagnac-type interferometer. (b) One photon of the entangled pair is sent to Bob, who will use it to reconstruct the quantum state using local unitaries. (c) The other photon of the pair is sent to Alice, where she prepares the path-encoded quantum state $\ket{\psi}$. This is followed by a polarization-path Bell-state measurement (BSM) using a PBS and HWP. Alice then sends the classical information to Bob, who reconstructs the state based on the received values.}
        \label{fig2}
\end{figure}

\twocolumngrid

Suppose, Alice wants to teleport an arbitrary qubit state represented as  
\begin{equation}  
\ket{\psi} = \alpha\ket{0} + \beta \ket{1},  
\end{equation}
to Bob where $\ket{0}$ and $\ket{1}$ denote the computational basis state and the complex coefficients $\alpha, \beta$ satisfying $|\alpha|^2 + |\beta|^2 = 1$.  To achieve this, she shares an entangled quantum channel and a classical communication channel with Bob. In order to teleport an arbitrary quantum state, Alice encodes the computational basis by introducing the spatial degree of freedom (DOF) to the one of the polarization entangled photon pair shared between her and Bob. Then, she performs a joint measurement known as the BSM involving single photon interferometer on a path qubit state and transmits the classical information based on her measurement outcome, allowing Bob to perform the necessary unitary operation and reconstruct the unknown quantum state on his side. The entire process of local operations and classical communication (LOCC) is carried out based on the coincidence measurement of the Alice and Bob photon pairs. This approach which Alice employs to teleport a quantum state by introducing path degree of freedom to the polarization entangled photon she shares, uses only entangled photons pair to represent three-qubit state and teleport an arbitrary quantum state. Since the scheme does not involve photon-photon interaction, the outcome will be deterministic.

In Fig.\,\ref{fig1}(a), the scheme for quantum teleportation of the path encoded qubit state using only entangled photons pair  in polarization degree of freedom shared between Alice and Bob is shown. The quantum state to be teleported is encoded on the entangled photon with Alice by introducing path degree of freedom. This is done by passing it through the variable beam splitter with splitting ratio corresponding the desired qubit state Alice wants to teleport to the Bob. This action of beam splitter is independent of the polarization state of the entangled photon Alice shares with Bob.  By performing the BSM on the polarization and path qubit state with Alice involving single photon interference on path qubit state, the measurement output from the two qubit state is obtained. The classical outcome of the measurement is communicated to the Bob and he will perform the corresponding unitary operator mapped to the classical information and reconstruct the qubit state on the polarization qubit with him. In Fig\,\ref{fig1}(b) the quantum circuit equivalent of the optical scheme is shown. The second and third qubit with Hadamard operation followed by controlled-NOT (C-NOT) operation represents the polarization entangled photon pair. The qubit state to teleport is encoded on Alice's photon via unitary operator ($U$) in the form of variable beam splitter.  The C-NOT operation, Hadamard operation followed by the joined measurement on the first two qubits represent the BSM on the path and polarization state of the photon with Alice. To perform Hadamard operation on the path qubit state one has to involve single photon interferometer\,\cite{KSKS2024}.

\begin{figure}[H]
        \centering
        \includegraphics[width = 0.5\textwidth]{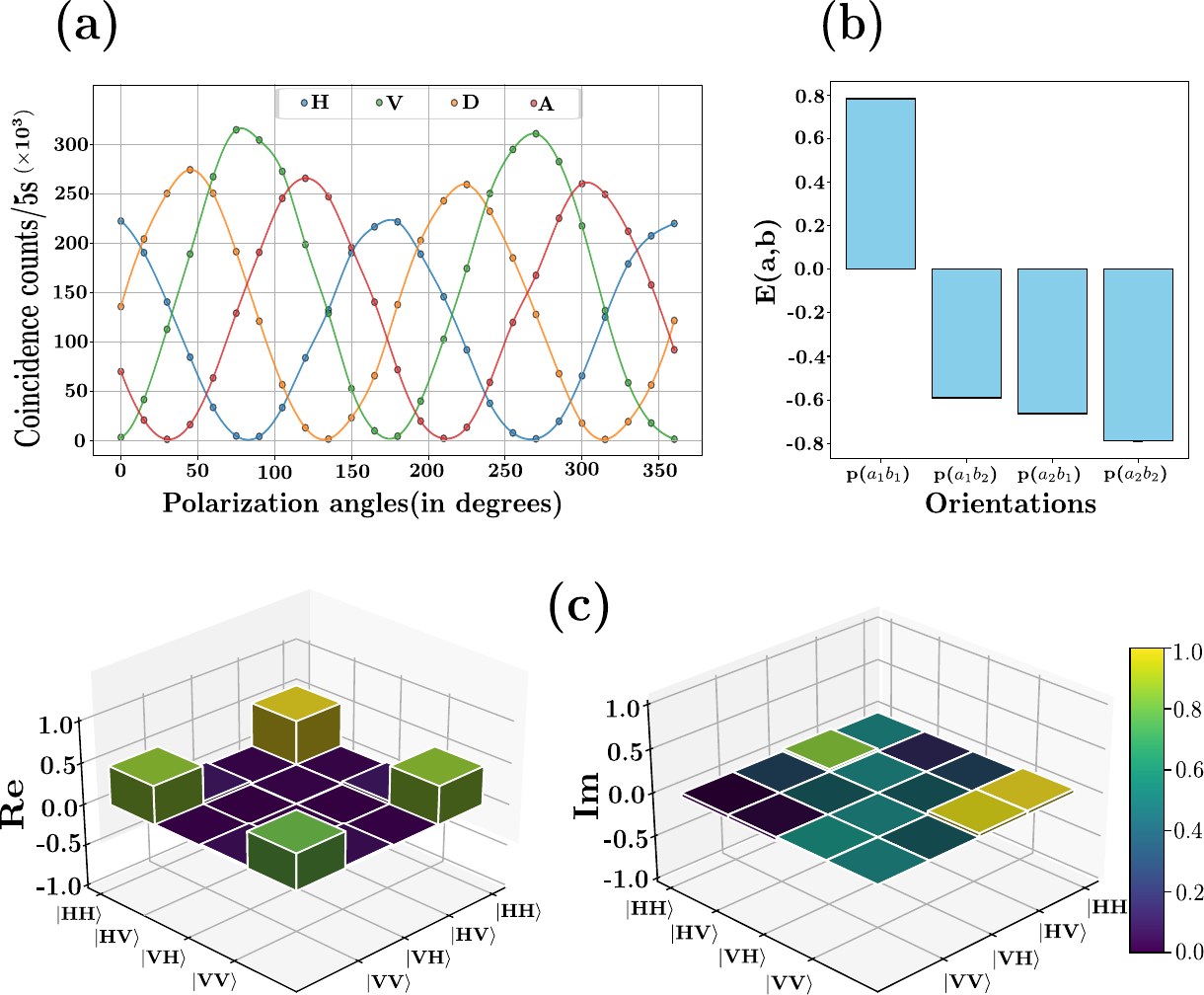}
        \caption{Entanglement characteristics of the SPDC source. In (a) the visibility in the four different basis is shown. The four different correlation values in the CHSH inequality is plotted in subfigure (b). The measured visibilities in the H/V and A/D bases were 98.9\%\ and 98.1\%\, respectively and a CHSH inequality violation of S = 2.72 $\pm$ 0.03. The quantum state tomography result in (c) shows that the entangled state is the bell state $\ket{\Phi^+}$.}
        \label{fig3}
\end{figure}

In Fig.\,\ref{fig2}, the complete experimental setup for quantum teleportation is shown. A polarization entangled photon pair AB represented by the state,
\begin{equation}
    \ket{\Phi^{+}}_{AB} = \frac{\ket{H}_{A}\ket{H}_B+\ket{V}_{A}\ket{V}_B}{\sqrt{2}}
\end{equation}
is prepared using a PPKTP crystal (Raicol-cyrstal), where quasi-phase matching is achieved through periodic poling with a period of $3.425\mu m$. This corresponds to type-0 phase matching, and we have utilized a Sagnac-loop configuration to generate the polarization entanglement\cite{MG2017,HOH2019}. Fig.\,\ref{fig2}(a) presented the entangled photon source generation unit, the results characterizing the entangled photon pair used for quantum teleportation is shown in Fig.\,\ref{fig3}. The measured visibilities in the H/V and A/D bases were 98.9\%\ and 98.1\%\, respectively [Fig.\,\ref{fig3}(a)]. The CHSH inequality violation, S = 2.72 $\pm$ 0.03 was obtained from the correlation values $E(a,b)$ shown in Fig.\,\ref{fig3}(b). The quantum state tomography results in Fig.\ref{fig3}(c) shows that the entangled state is the Bell state $\ket{\Phi^+}$. 

One of the entangled photon pair A is sent to Alice for encoding the qubit state to be teleported and the other pair B is sent to Bob for state reconstruction.  To encode the arbitrary qubit state to be teleported, the photon A is passed through a variable beam-splitter (BS) whose output will be the path encoded qubit state. 
Generally, a beam splitter matrix $B(\theta,\phi)$ takes the form:  
$$ U_{\text{BS}}(\theta,\phi) = e^{i\phi/2}  
\begin{bmatrix}  
\cos\theta & i\sin\theta \\  
i\sin\theta & \cos\theta  
\end{bmatrix}. $$
This matrix describes how the creation and annihilation operators transform under the beam splitter. If we consider the input mode as $\ket{0}$, it transforms as follows:
$\ket{\psi}_{\text{out}} = \cos(\theta)\ket{0} + i\sin(\theta)\ket{1}$, up to a global phase. Now, if we introduce a relative phase between the spatial modes, a general path-encoded quantum state takes the form: 
$$\ket{\psi}_{\text{out}} = \cos(\theta)\ket{0} + i e^{i\phi} \sin(\theta)\ket{1}.$$ This will also be the state we want to teleport to Bob. For simplicity, we write it as  
\begin{equation}  
\ket{\psi} = \alpha\ket{0} + \beta \ket{1} = \sqrt{\eta}\ket{0} + e^{i\phi}\sqrt{1-\eta}\ket{1},  
\end{equation}  
such that $|\alpha|^2 + |\beta|^2 = 1$, where $\eta$ and $(1-\eta)$ represents reflection and transmission probability, respectively. This transformation does not affect other degrees of freedom of the photon, such as polarization in this case.
A non-polarizing beam-splitter with different splitting ratios has been utilized to produce these quantum state. However, this operation is restricted due to the limitation on the availability of BS with desired splitting ratio and in addition to  
difference in splitting for S and P polarization vowing to the manufacturing limitations.  However, as shown in Fig.\,\ref{fig2} Alice's unit, using two polarizing beam-splitters (PBS) and two half-wave plates in combination, a variable beam-splitter with different ratios can be created\,\cite{YFWG2019,RJXT2016}. The path length is matched within the coherence length of the photons, and the phase is locked using a piezo to stabilize the path-encoded state.

However, before introducing the arbitrary path state, the path encoding has to be  added on Alice's side, resulting in the state taking the form,
\begin{equation}
    \ket{\Psi} = \frac{1}{\sqrt{2}}\big(\ket{H}_{A}\ket{0}_{AP}\ket{H}_B+\ket{V}_{A}\ket{0}_{AP}\ket{V}_B\big)
\end{equation}
where \(\ket{0}_{AP}\) represents the spatial mode of the photon before the path-encoded quantum state is prepared by Alice. As mentioned earlier, Alice can encode arbitrary qubit state by using a variable beam-splitter along the path of the photon with her, The complete quantum state after path coding is given by,
\begin{equation}
    \begin{aligned}
        \ket{\Psi}_{1} = \frac{1}{\sqrt{2}} \Big( & \ket{H}_{A}(\alpha\ket{0}_{AP}+\beta\ket{1}_{AP})\ket{H}_B \\
        & + \ket{V}_{A}(\alpha\ket{0}_{AP}+\beta\ket{1}_{AP})\ket{V}_B \Big).
    \end{aligned}
\end{equation}
In the preceding expression we can see that the three-qubit state suitable for quantum teleportation of one of the qubit state using only two entangled photons in polarization degree and a path degree of freedom. 

To teleport a quantum state BSM is performed on the qubits with Alice. The first step of BSM is performed by using a PBS which acts as a C-NOT operation with polarization as the control and path as the target. This transforms the state $\ket{\Psi}$ to:
\begin{equation}
    \begin{aligned}
        \ket{\Psi}_{2} = \frac{1}{\sqrt{2}} \Big( & \ket{H}_{A}(\alpha\ket{0}_{AP}+\beta\ket{1}_{AP})\ket{H}_B \\
        & + \ket{V}_{A}(\alpha\ket{1}_{AP}+\beta\ket{0}_{AP})\ket{V}_B \Big).
    \end{aligned}
\end{equation}

Then a Half-wave plate (HWP) is used to realize the Hadamard operation on the path qubit state. The Hadamard operation should produce a coherent superposition. It is achieved by allowing the single photon from two paths to interfere using a single photon interference with high visibility. A single photon interference with high visibility is required to maintain the purity of teleported states.  We achieved a maximum of 83\% visibility. After the complete BSM the final state will be, 
\begin{equation}
    \begin{aligned}
        \ket{\Psi}_{F} = \frac{1}{\sqrt{2}} \Big( & \ket{H}_{A}\ket{0}_{AP}(\alpha\ket{H}_B+\beta\ket{V}_B) + \\
        & \ket{V}_{A}\ket{0}_{AP}(\alpha\ket{H}_B-\beta\ket{V}_B)+ \\
        & \ket{H}_{A}\ket{1}_{AP}(\beta\ket{H}_B+\alpha\ket{V}_B)+ \\
        &\ket{V}_{A}\ket{1}_{AP}(\beta\ket{H}_B-\alpha\ket{V}_B)\Big). \\
    \end{aligned}
\end{equation}

\begin{figure}[H]\centering
        \includegraphics[width = 0.5\textwidth]{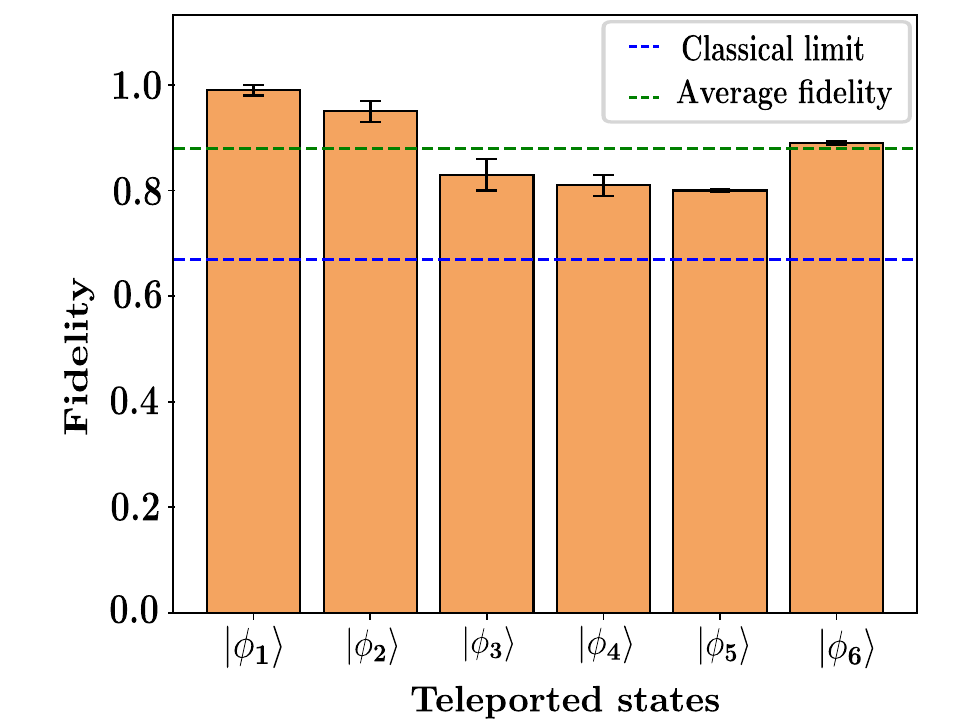}
        \caption{Experimental results for quantum teleportation. We considered the coincidence counts between the detectors in Alice's side and Bob's side to obtain the stokes parameters and reconstructed the polarization state of the Bob's photon. Six states: $\ket{\phi_1}=\ket{0}, \ket{\phi_2}=\ket{1}, \ket{\phi_3}=(\ket{0}+\ket{1})/\sqrt{2}, \ket{\phi_4}=(\ket{0}-\ket{1})/\sqrt{2}, \ket{\phi_5}= (\ket{0}+i\ket{1})/\sqrt{2}, \ket{\phi_6}=(\ket{0}-i\ket{1})/\sqrt{2}$ were encoded and teleported over a distance of 2 meters with an average fidelity of F = 0.88$\pm$0.04.}
        \label{fig4}
\end{figure}

To perform the measurement of the output state,  PBS is used and four different states are channeled to four different detectors. The coincidence measurement between any one of the photon A (with Alice) and B  (with Bob) is utilized to get the classical output in the form of two classical bit. These  bits are classical communicated to Bob who in turn uses appropriate unitary operation on the photon on his side to reconstruct the qubit state which Alice encoded on path degree of freedom in the polarization qubit with him.

\begin{figure}[H]
        \centering
        \includegraphics[width = 0.5\textwidth]{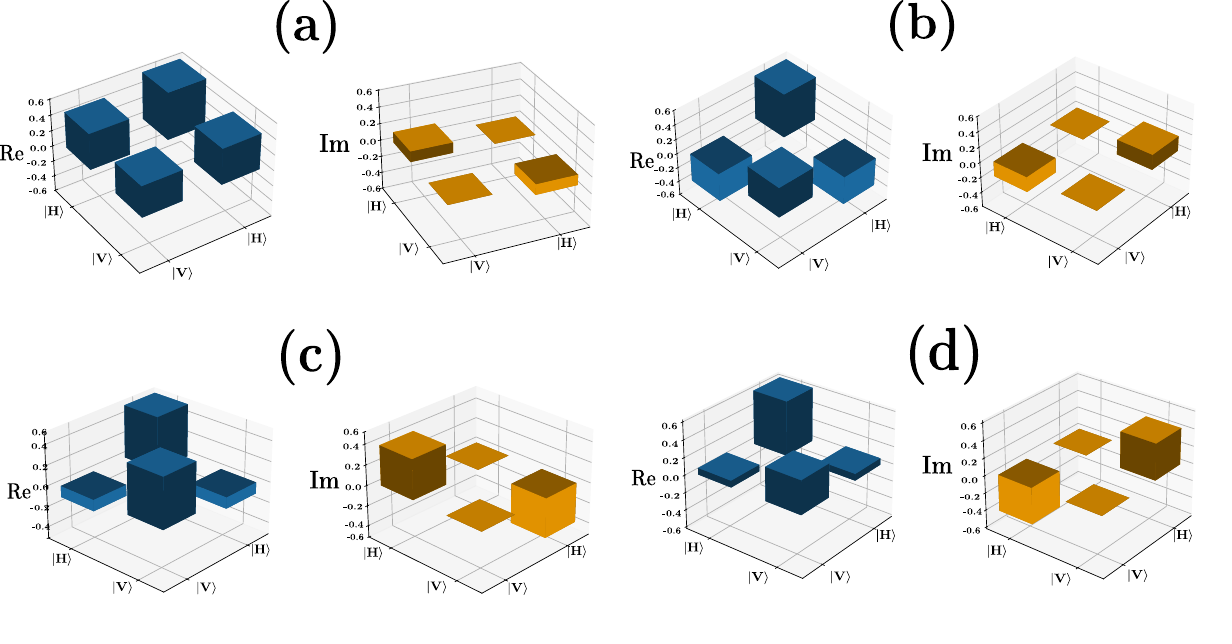}
        \caption{Tomography of the teleported states: $|\phi_3\rangle, |\phi_4\rangle, |\phi_5\rangle, |\phi_6\rangle$. Quantum state tomography was done for 4 qubit states using the stoke's parameter obtained from the coincidence measurements. Fig (a), (b), (c), (d) shows the real and imaginary parts of the density matrix of $|+\rangle, |-\rangle, |R\rangle, |L\rangle$, respectively.}
        \label{fig5}
\end{figure}

\begin{figure}[H]
        \centering
        \includegraphics[width = 0.5\textwidth]{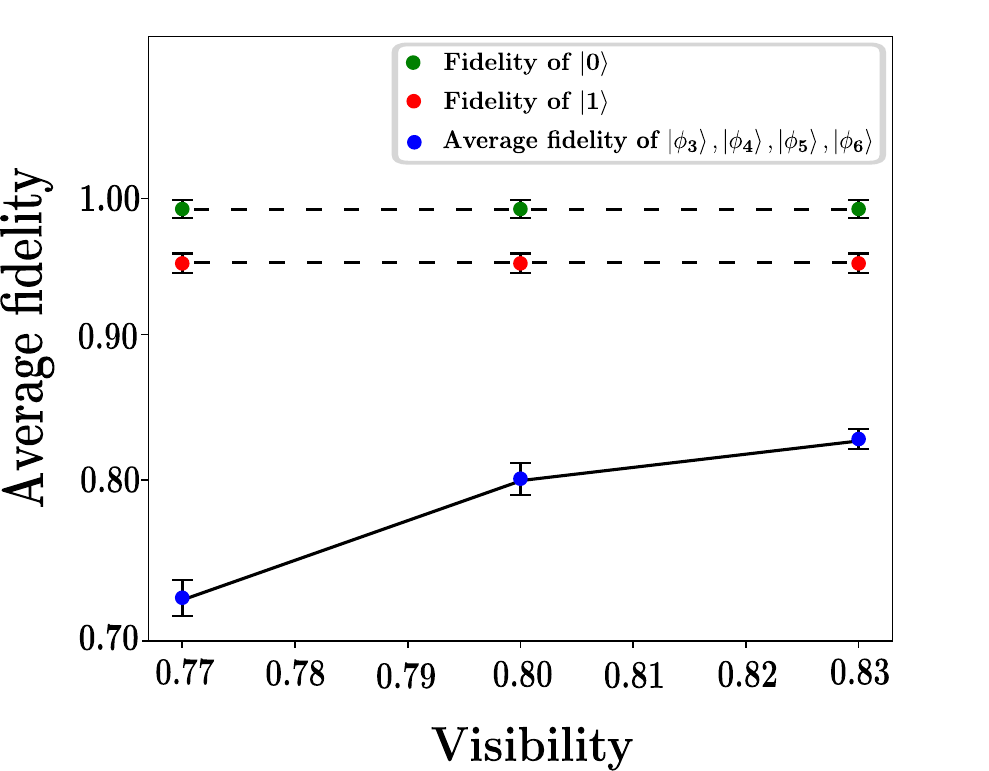}
        \caption{Single photon interference visibility v/s Average fidelity of the teleported states.  Three different visibilities were obtained by adjusting the piezo controller and the fidelities were calculated for all the six states. The average fidelity of $|0\rangle, |1\rangle$ remained constant since no interference is needed. But the fidelities of $|+\rangle, |-\rangle, |R\rangle, |L\rangle$ increases as the single photon interference visibility increases. This qualitative analysis is in accordance with the theoretical predictions.}
        \label{fig6}
\end{figure}

We prepared and teleported 6 different state: $\ket{\phi_1}=\ket{0}, \ket{\phi_2}=\ket{1}, \ket{\phi_3}=(\ket{0}+\ket{1})/\sqrt{2}, \ket{\phi_4}=(\ket{0}-\ket{1})/\sqrt{2}, \ket{\phi_5}= (\ket{0}+i\ket{1})/\sqrt{2}, \ket{\phi_6}=(\ket{0}-i\ket{1})/\sqrt{2}$. Fig.\,\ref{fig4} shows the fidelity of all the 6 states with an average fidelity F = 0.88$\pm$0.04 which is greater than the classical bound of 0.67.  Tomography of the teleported four different qubit states is also shown in Fig.\,\ref{fig5}
The  highest fidelity we have obtained is 99.00\% for teleporting a state $|0\rangle$ and lowest of 80.30\% for state $|R\rangle = (|0\rangle + i |1\rangle)/\sqrt{2})$. In Fig.\ref{fig6}, Fidelity of teleported state for different single photon interference visibility is shown. This shows that the fidelity of teleportation is highly dependent on the interference visibility of the single photon interference employed for BSM in the scheme. The average fidelity of $|0\rangle$ and $|1\rangle$ remains constant since no interference is needed for BSM on them. However, for the state $|1\rangle$, an additional reflection in the path caused reduced coupling, resulting in a lower fidelity compared to the state $|0\rangle$.
The experiment performed on a tabletop setup  had Alice and Bob separated by a distance of 2 meters from each other.  With this we can establish that by increase in single photon interference visibility near cent percent teleportation can be achieved. This ensures a robust and deterministic non-classical teleportation. 

{\it Conclusion :}
In summary we have reported the deterministic teleportation of qubit state with high fidelity  than the other qubit teleportation protocols reported in photonic systems.  The possibility of encoding any arbitrary qubit state into path degree of freedom of the entangled photon pair makes the scheme easily employable for practical applications.  The role of single-photon interference in the path qubit state is crucial for achieving high teleportation fidelity. Notably, except for  teleporting the states $\ket{0}$ or $\ket{1}$, the interference in the path qubit state directly influences the fidelity of the teleported state. By employing high precision single photon interferometer on path qubit state for BSM, a near cent percent fidelity can be achieved.  Though the experiment was performed  with only a separation of two meter distance from Alice and Bob, the distance will only be limited by the ability to share the entangled photons between two parties.  The  scheme proposed can be adopted to  encoding qubit states into other degrees of freedom of photons and realise deterministic teleportation. This can also be employed for  teleportation of higher-dimensional quantum states and increase the probability to teleportation. The demonstrated high fidelity, combined with the deterministic nature of our scheme, opens up new possibilities for practical quantum communication, quantum network and quantum computation applications.


\end{document}